\documentclass[
    ,final            % use final for the camera ready runs
  ]
  {aipproc}

\layoutstyle{8x11single}

\def\be{\begin{equation}}
\def\ee{\end{equation}}
\def\bea{\begin{eqnarray}}
\def\eea{\end{eqnarray}}
\def\bear{\begin{array}}
\def\ear{\end{array}}
\def\bfig{\begin{figure}}
\def\efig{\end{figure}}
\def\bcen{\begin{center}}
\def\ecen{\end{center}}
\def\bi{\begin{itemize}}
\def\ei{\end{itemize}}

\def\raw{\rightarrow}

\def\slash{\!\!\!\! /}
\begin{document}

\title{Photon emission in neutral current interactions with nucleons and nuclei}


\author{L. Alvarez-Ruso}{
  address={Instituto de F\'isica Corpuscular and Departamento de F\'\i sica Te\'orica , \\Centro Mixto
Universidad de Valencia-CSIC, E-46071 Valencia, Spain}
}

\author{J. Nieves}{
  address={Instituto de F\'isica Corpuscular and Departamento de F\'\i sica Te\'orica , \\Centro Mixto
Universidad de Valencia-CSIC, E-46071 Valencia, Spain}
}

\author{E. Wang}{
  address={Instituto de F\'isica Corpuscular and Departamento de F\'\i sica Te\'orica , \\Centro Mixto
Universidad de Valencia-CSIC, E-46071 Valencia, Spain} 
}

\begin{abstract}
We report on our study of photon emission induced by $E_\nu \sim 1$~GeV (anti)neutrino neutral current interactions with nucleons and nuclei. This process is an important background for $\nu_e$ appearance oscillation experiments. At the relevant energies, the reaction is dominated by the excitation of the $\Delta (1232)$ resonance but there are also non-resonant contributions that, close to threshold, are fully determined by the effective chiral Lagrangian of strong interactions. We have obtained differential and integrated cross section for the (anti)neutrino-nucleon scattering and compare them with previous results. Furthermore, we have extended the model to nuclear targets taking into account Fermi motion, Pauli blocking and the in-medium modifications of the $\Delta$ properties. This study is important in order to reduce systematic effects in neutrino oscillation experiments.
\end{abstract}

\maketitle

%%%%%%%%%%%%%%%%%%%%%%%%%%%%%%%%%%%%%%%%%%%%
%% MAINMATTER
%%%%%%%%%%%%%%%%%%%%%%%%%%%%%%%%%%%%%%%%%%%%

\section{Introduction}

A good understanding of (anti)neutrino interactions with nuclear targets in the $E_\nu \sim 1$~GeV region is vital to reduce the systematic uncertainties in oscillation experiments aiming at a precise determination of neutrino properties. One of the possible reaction channels is photon emission induced by neutral current interactions (NC$\gamma$), which can occur on single nucleons
\be
\label{eq:reac_nucleon}
\nu (\bar{\nu})\, N \rightarrow \nu (\bar{\nu})\, \gamma \, N \,,
\ee
and on nuclear targets
\bea
\label{eq:reac_incoh}
\nu (\bar{\nu})\, A &\rightarrow& \nu (\bar{\nu})\, \gamma \, X  \, \\
\label{eq:reac_coh}
\nu (\bar{\nu})\, A &\rightarrow& \nu (\bar{\nu})\, \gamma \, A \,\\
\nu (\bar{\nu})\, A &\rightarrow& \nu (\bar{\nu})\, A'^* \, N \label{eq:reac_2step} \nonumber \\
&\raw&  \nu (\bar{\nu}) \, \gamma \, A' \, N \,,
\eea
with incoherent [Eq.~(\ref{eq:reac_incoh})] or coherent  [Eq.~(\ref{eq:reac_coh})] reaction mechanisms. It is also possible that, after nucleon knockout, the residual excited nucleus decays emitting $\gamma$ rays. This mechanism has been recently investigated~\cite{Ankowski:2011ei} and shall not be discussed here. 

Weak photon emission has a small cross section compared, for example with pion production (NC$\pi$), the most important inelastic mechanism. Indeed, while NC$\pi$ takes place predominantly via a weak interaction followed by a strong decay, in the case of NC$\gamma$ one has a much weaker electromagnetic vertex instead of the strong one. In spite of this, NC$\gamma$ turns out to be one of the largest backgrounds in $\nu_\mu \rightarrow \nu_e$($\bar{\nu}_\mu \rightarrow \bar{\nu}_e$) experiments when $\gamma$'s are misidentified as $e^\mp$ from charge-current quasi-elastic scattering of $\nu_e (\bar{\nu}_e)$. 

This is precisely the situation in the MiniBooNE experiment, where the gamma background is estimated from the measured NC$\pi^0$ rate assuming that it comes form radiative decay of weakly produced resonances, mainly $\Delta \raw N \, \gamma$. The experiment finds an excess of events with respect to the predicted background in both $\nu $ and $\bar{\nu}$ modes. In the $\bar{\nu}$ mode, the data are found to be consistent with $\bar{\nu}_\mu \raw \bar{\nu}_e$ oscillations and have some overlap with the LSND result~\cite{Aguilar-Arevalo:2013pmq}. In contrast, the reconstructed energy distribution of the $e$-like events in the $\nu$ mode is only marginally compatible with  a simple two-neutrino oscillation model, exhibiting an unexplained excess of events for $E_\nu^{QE} < 475$~MeV~\cite{Aguilar-Arevalo:2013pmq,AguilarArevalo:2008rc}. While several exotic explanations for this excess have been proposed, it could be related to unknown systematics and backgrounds. In the kinematic region where this anomaly is observed, NC$\gamma$ is the second largest background behind NC$\pi^0$. In view of this, it is important to  study this process in detail, using the well developed framework of hadronic and nuclear physics.           

The first step in this direction was performed in Ref~\cite{Hill:2009ek}, where reaction~(\ref{eq:reac_nucleon}) was studied with a microscopic model developed in terms of hadronic degrees of freedom: nucleon, $\Delta(1232)$ resonance and mesons. With this model, the NC$\gamma$ events at MiniBooNE were calculated to be twice as many as expected from the MiniBooNE {\it in situ} estimate. The conclusion was that NC$\gamma$ events give a significant contribution to the low-energy excess of $e$-like events~\cite{Hill:2010zy}. However, in Ref.~\cite{Hill:2010zy} the nuclear target ($^{12}C$) was treated as an ensemble of nucleons, neglecting all nuclear-medium corrections. Furthermore, an energy independent and rather high efficiency correction compared with the presently available figures~\cite{MiniBooNEweb} was assumed in the analysis. A contrasting result, much closer to the MiniBooNE estimate, was recently obtained in Ref~\cite{Zhang:2012xn}, based on the chiral effective field theory of nuclei~\cite{Zhang:2012aka}, phenomenologically extended to the intermediate energies ($E_\nu \sim 1$~GeV) of the $\nu$ flux at MiniBooNE. Our approach, described in the next section, has several ingredients in common with these previous works but also quantitative differences.   
  
\section{Formalism}

\begin{figure}[t]
\label{fig:diags}
\includegraphics[width=0.19\textwidth]{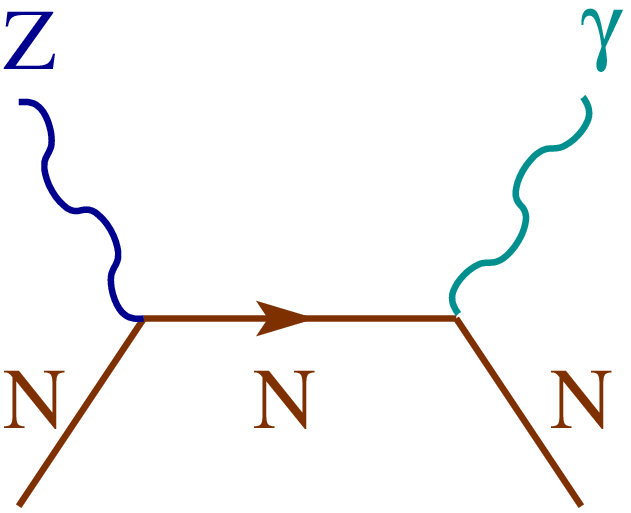}
\hspace{.02\textwidth} 
\includegraphics[width=0.19\textwidth]{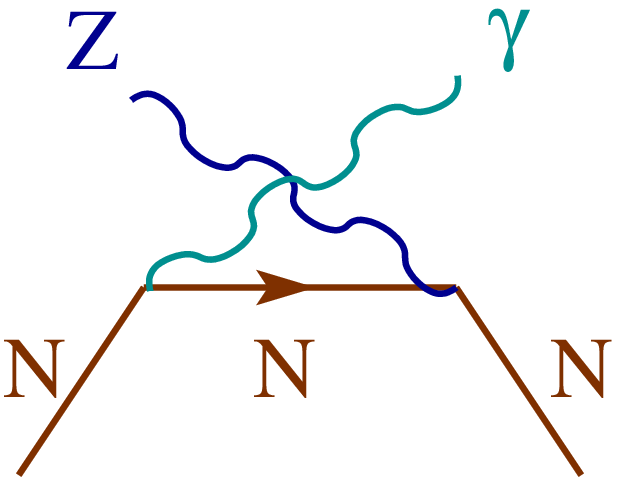}
\hspace{.02\textwidth} 
\includegraphics[width=0.19\textwidth]{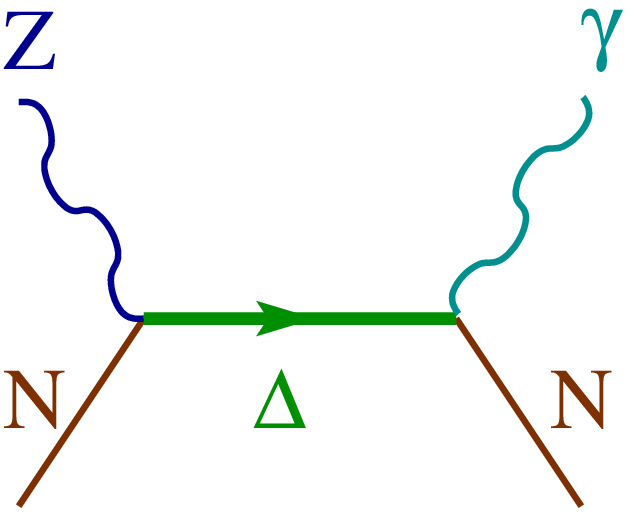}
\hspace{.02\textwidth}  
\includegraphics[width=0.19\textwidth]{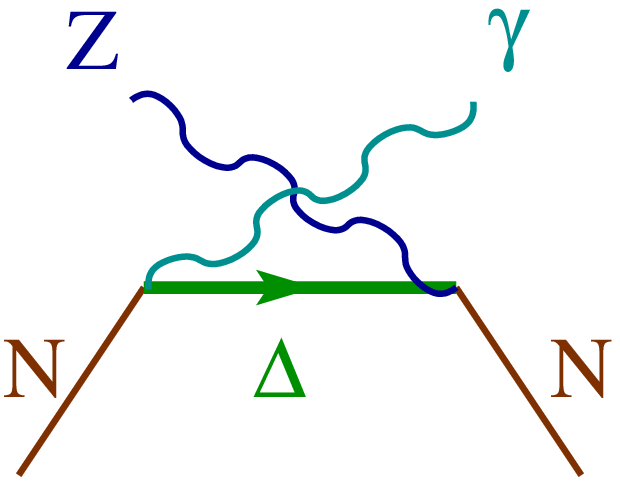} 
\hspace{.02\textwidth} 
\includegraphics[width=0.16\textwidth]{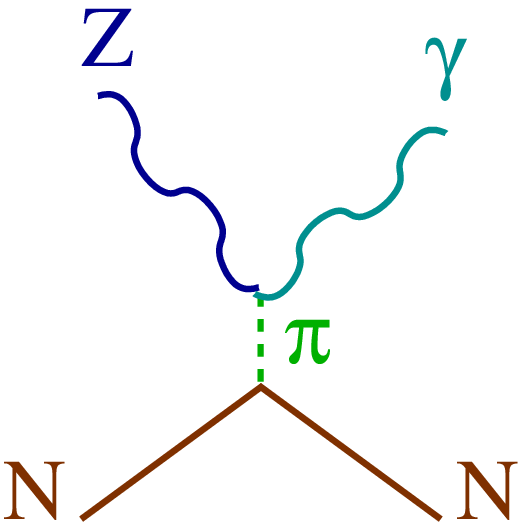} 
\caption{Set of Feynman diagrams of the model}
\end{figure}

First, let us discuss NC$\gamma$ on the nucleon. The amplitude can be cast as 
\be
{\mathcal M}_r =  \frac{G_F e}{\sqrt{2}} \epsilon^{*(r)}_\mu \bar{u}(p') \Gamma^{\mu\alpha} u(p)\, l_\alpha \,.
\ee
It is proportional to the Fermi constant $G_F$  and the electric charge $e$; $\epsilon^{*(r)}$ denotes the photon polarization vector; $l_\alpha$, the Standard Model neutral current for (anti)neutrinos, is contracted with the hadronic current $\bar{u} \Gamma^{\mu\alpha} u$, which is specific for the different reaction mechanisms. The later is determined by the set of Feynman diagrams shown in Fig.~\ref{fig:diags}: direct and crossed nucleon-pole and $\Delta$-pole terms, and pion-pole term 
\be
\Gamma^{\mu\alpha} = \Gamma^{\mu\alpha}_N+\Gamma^{\mu\alpha}_{\Delta}+\Gamma^{\mu\alpha}_{\pi} \,.
\ee 
For the nucleon-pole diagrams, $\Gamma^{\mu\alpha}_N$ takes the form
\be
\Gamma^{\mu\alpha}_N = J^\mu_{EM}(-q_\gamma) D_N(p+q) J^\alpha_{NC}(q) + J^\alpha_{NC}(q) D_N(p'-q)  J^\mu_{EM}(-q_\gamma)
\ee
where $D_N(p) = (p \slash - M)^{-1}$ is the nucleon propagator; $q$ is the 4-momentum transferred to the nucleon and $q_\gamma$, the one of the outgoing photon.
\begin{eqnarray}
J^\alpha_{NC}(q)&=&\gamma^\alpha \tilde{F}_1(q^2)+\frac{i}{2M}\sigma^{\alpha\beta}q_\beta \tilde{F}_2(q^2) - \gamma^\alpha \gamma_5 \tilde{F}_A(q^2),  \nonumber \\
J^\mu_{EM}(-q_\gamma)&=& \gamma^\mu F_1(0) - \frac{i}{2M}\sigma^{\mu\nu}q_{\gamma\nu} F_2(0),
\end{eqnarray}
where $\tilde{F}_{1,2}$ and $F_{1,2}$ are the vector NC and EM form factors, respectively. For the axial form factor
\be
2 \tilde{F}^{(p,n)}_A  =  \pm F_A  +F^{(s)}_A
\ee 
%\,, \qquad
we take
\be
F_A(Q^2) = g_A \left(1-\frac{q^2}{M^2_A} \right)^{-2}
\ee
with $g_A=1.267$, the axial coupling, and $M_A=1.016$~GeV~\cite{Bodek:2007ym}, neglecting the strange part of the axial form factor $F^{(s)}_A$, as well as in the vector NC ones. 

The most prominent contribution to the cross section arises from the weak excitation of the $\Delta(1232)$ resonance followed by its radiative decay. For direct and crossed $\Delta$ terms one has
\be
\Gamma^{\mu\alpha} = \tilde{J}^{\delta\mu}_{EM}(p^\prime,q_\gamma) D^\Delta_{\delta\sigma}(p+q) J^{\sigma\alpha}_{NC}(p,q)
+ \tilde{J}^{\delta\alpha}_{NC}(p^\prime,-q)  D^\Delta_{\delta\sigma}(p'-q) J^{\sigma\mu}_{EM}(p,-q_\gamma) \,,
\ee 
where $\tilde{J}^{\alpha\beta} = \gamma_0\, \left( J^{\alpha\beta} \right)^\dagger \, \gamma_0$ and the $\Delta$ propagator is given by
\be
\label{eq:Deltaprop}
D^\Delta_{\delta\sigma} (p) = \frac{\Lambda_{\delta\sigma}}{p^2-M^2_\Delta+i M_\Delta\Gamma_\Delta(p^2)} \,;
\ee  
$\Lambda_{\delta\sigma}$ is the $N-\Delta$ projector and $\Gamma_\Delta(p^2)$ the $\Delta$ (energy dependent) width dominated by the $\Delta \raw N \, \pi$ $p$-wave decay. The vertices $J^{\beta\mu}_{NC,EM}$ can be written in the most general form as 
\begin{eqnarray}
J^{\beta\mu}_{NC}(p,q) &=& \left[ \frac{\tilde{C}^V_3(q^2)}{M} (g^{\beta\mu} q \slash
    -q^\beta \gamma^\mu ) +\frac{\tilde{C}^V_4(q^2)}{M^2} (g^{\beta\mu} q \cdot p_\Delta
    -q^\beta p^\mu_\Delta )+\frac{\tilde{C}^V_5(q^2)}{M^2} (g^{\beta\mu} q \cdot p
    -q^\beta p^\mu ) \right] \gamma_5, \nonumber \\
 &+&\frac{\tilde{C}^A_3(q^2)}{M} (g^{\beta\mu} q \slash -q^\beta \gamma^\mu ) 
   + \frac{\tilde{C}^A_4(q^2)}{M^2} (g^{\beta\mu} q \cdot p_\Delta
    -q^\beta p^\mu_\Delta ) + \frac{\tilde{C}^A_5(q^2)}{M^2} g^{\beta\mu} \,,
\end{eqnarray}
\be
J^{\beta\mu}_{EM}(p,q_\gamma) = \left[ \frac{C^{EM}_3(0)}{M} (g^{\beta\mu} q\slash_\gamma  -q_\gamma^{\prime\beta} \gamma^\mu ) +\frac{C^{EM}_4(0)}{M^2} (g^{\beta\mu} q_\gamma \cdot p_{\Delta} -q_\gamma^\beta p^\mu_{\Delta} )+\frac{C^{EM}_5(0)}{M^2} (g^{\beta\mu} q_\gamma \cdot p -q_\gamma^\beta p^\mu ) \right] \gamma_5 \,,
\ee
in terms of the EM vector and NC vector and axial $N-\Delta$ transition form factors $C^{EM}_i =-C^V_i$, $\tilde{C}^V_i=-(1-2 \sin^2{\theta_W}) C^V_i$ and  $\tilde{C}^A_i=-C^A_i$, respectively; $C^{EM}_i$ can be related to the $N-\Delta(1232)$ helicity amplitudes, for which, following Ref.~\cite{Leitner:2008ue}, we adopt the parametrizations obtained by the global analysis of $\pi$ photo- and electro-production data with the unitary isobar model MAID~\cite{Drechsel:2007if}. In the axial sector we assume $\tilde{C}^A_3 = 0$ and $\tilde{C}^A_4 = -C^A_5/4$ for the subleading (in a $q^2$ expansion) form factors, while for the dominant $C^A_5$ we take
\be
C^A_5(q^2) = C^A_5(0) \left(1-\frac{q^2}{M^2_{A \Delta}} \right)^{-2}\,,
\ee
with $C^A_5(0)=1.0\pm0.11$ and $M_A=0.93 \pm 0.07$~GeV fixed in a fit to $\nu_\mu \, d \raw \mu^- \Delta^{++} \, n$ BNL and ANL data~\cite{Hernandez:2010bx}. The error in  $C^A_5(0)$ is the main source of theoretical uncertainty in our predictions. 

The first four diagrams in Fig~\ref{fig:diags} account for the leading contributions close to threshold. The inclusion of form factors allow to extend the model to the higher energies of interest for neutrino experiments. The last, pion-pole, term is of higher order and should be small. It is indeed found to be negligible compared to the mechanisms discussed above. We assume that other higher order terms can be also neglected.     

The model for NC$\gamma$ on the nucleon has been extended to the incoherent reaction (\ref{eq:reac_incoh}) on nuclear targets.  For this purpose we have adopted the relativistic local Fermi gas description, according to which the target nucleons have momenta up to a Fermi momentum defined locally $p^{p,n}_F(\vec{r})=\left[3 \pi^2\rho_{p,n}(\vec{r})\right]^{1/3}$ as a function of the local density of protons and neutrons independently. The density distributions are based on empirical determinations in the case of protons and on realistic theoretical models in the case of neutrons. Final nucleons are not allowed to take occupied states (Pauli blocking).

Furthermore, it is known that the properties of the $\Delta(1232)$ resonance get substantially modified in a nuclear environment. This nuclear effect can be taken into account by performing the following substitutions in the $\Delta(1232)$ propagator [Eq.~(\ref{eq:Deltaprop})] 
\bea
M_\Delta &\raw& M_\Delta + \mathrm{Re} \Sigma_\Delta(\rho)\,, \\
\Gamma_\Delta &\raw& \tilde{\Gamma}_\Delta - 2\,  \mathrm{Im} \Sigma_\Delta(\rho)\,.
\eea
The real part of the in-medium $\Delta$ selfenergy, $\Sigma_\Delta$, receives an attractive (negative) contribution from the nuclear mean field, which is partially cancelled by an effective repulsive piece from iterated $\Delta$-hole excitations. In view of this and for the sake of simplicity we take $\mathrm{Re} \Sigma_\Delta(\rho) \approx 0$. The resonance decay width is reduced to  $\tilde{\Gamma}_\Delta$ because the final nucleon in $\Delta \raw \pi N$ can be Pauli blocked but, on the other hand, it increases because of the presence of many body processes such as $\Delta \, N \raw N \, N$, $\Delta \, N \raw N \,N \, \pi$ and  $\Delta \, N \, N \raw N \, N \, N$ (collisional broadening). These new decay channels, which are accounted in $\mathrm{Im} \Sigma_\Delta$, have been parametrized as a function of the local density in Ref.~\cite{Oset:1987re}. 

\section{Results}

 \begin{figure}[ht]
\label{fig:resu_nucleon}
\includegraphics[width=0.5\textwidth]{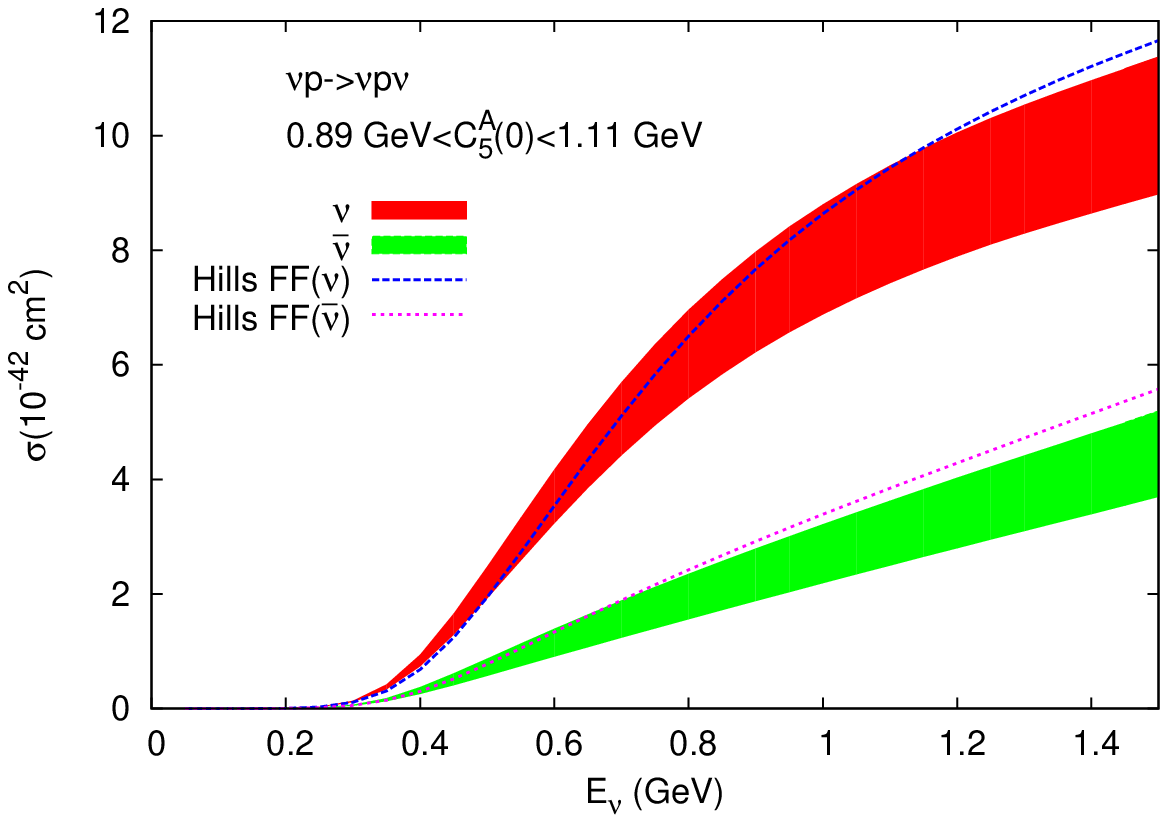}
\includegraphics[width=0.5\textwidth]{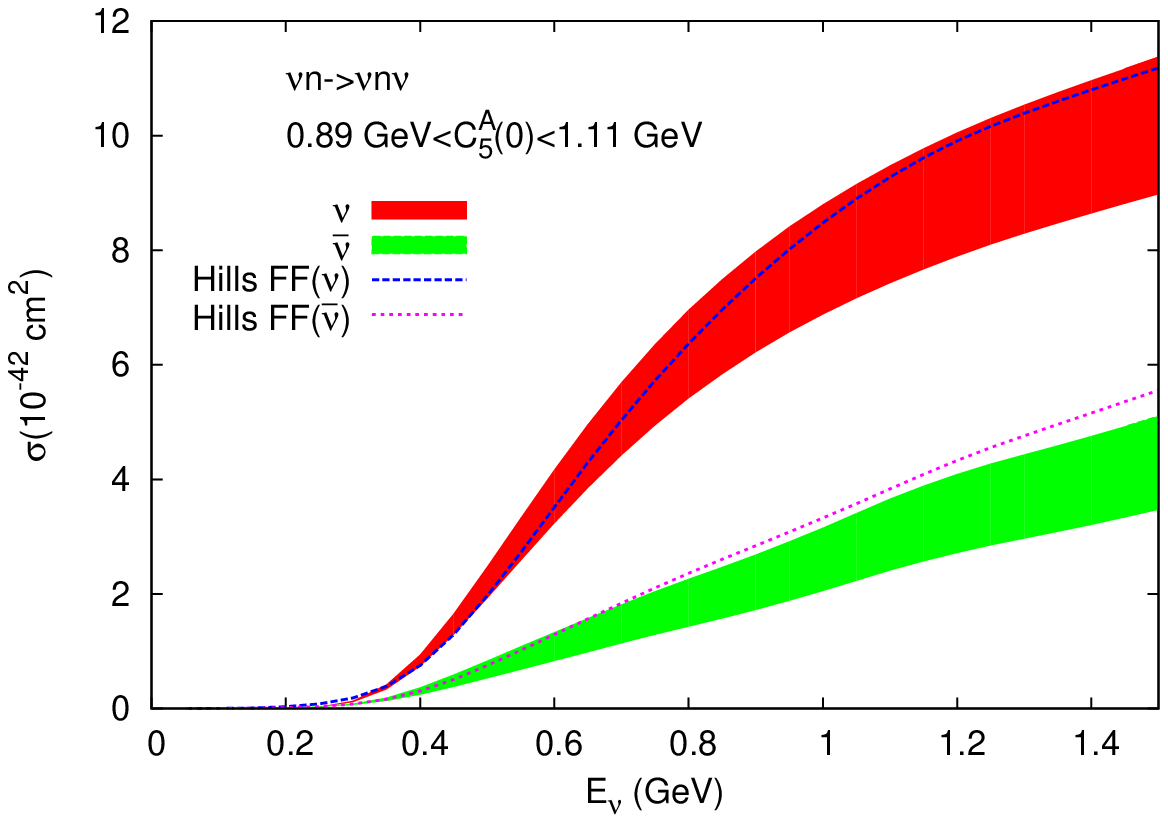}
  \caption{Integrated cross section for $(\bar{\nu})\nu \, p \raw (\bar{\nu})\nu \, p \, \gamma$ (left panel) and $(\bar{\nu})\nu \, n \raw (\bar{\nu})\nu \, n \, \gamma$ (right panel).}
\end{figure}
We consider first NC$\gamma$ on single nucleons. The integrated cross sections as a function of the (anti)neutrino energies are given in Fig.~\ref{fig:resu_nucleon}. The $\Delta$ mechanism is dominant but, at $E_\nu \sim 1.5$~GeV, the cross section from nucleon-pole terms is only $\sim 2.5$ smaller than the $\Delta$ one. The error bands are determined by the uncertainty in $C^A_5(0)$ discussed above. As in other weak interaction processes, the different helicities of $\nu$ and $\bar{\nu}$ are responsible for different interferences, resulting in smaller $\bar{\nu}$ cross sections with a more linear energy dependence. The dashed and dotted curves are obtained with the assumptions of Ref.~\cite{Hill:2009ek}. The small differences, mainly in the faster rise of the cross section can be explained by:  a larger $C^A_5(0)=1.2$ {\it vs} $1\pm0.11$ of the present work, a constant $\Gamma_\Delta =120$~MeV of Ref,~\cite{Hill:2009ek} {\it vs} the energy dependent width used in this work, and an $M_A =1.2$~GeV {\it vs} 1~GeV taken here for the nucleon-pole terms.   

\begin{figure}[hb]
\label{fig:resu_carbon}
\includegraphics[width=0.5\textwidth]{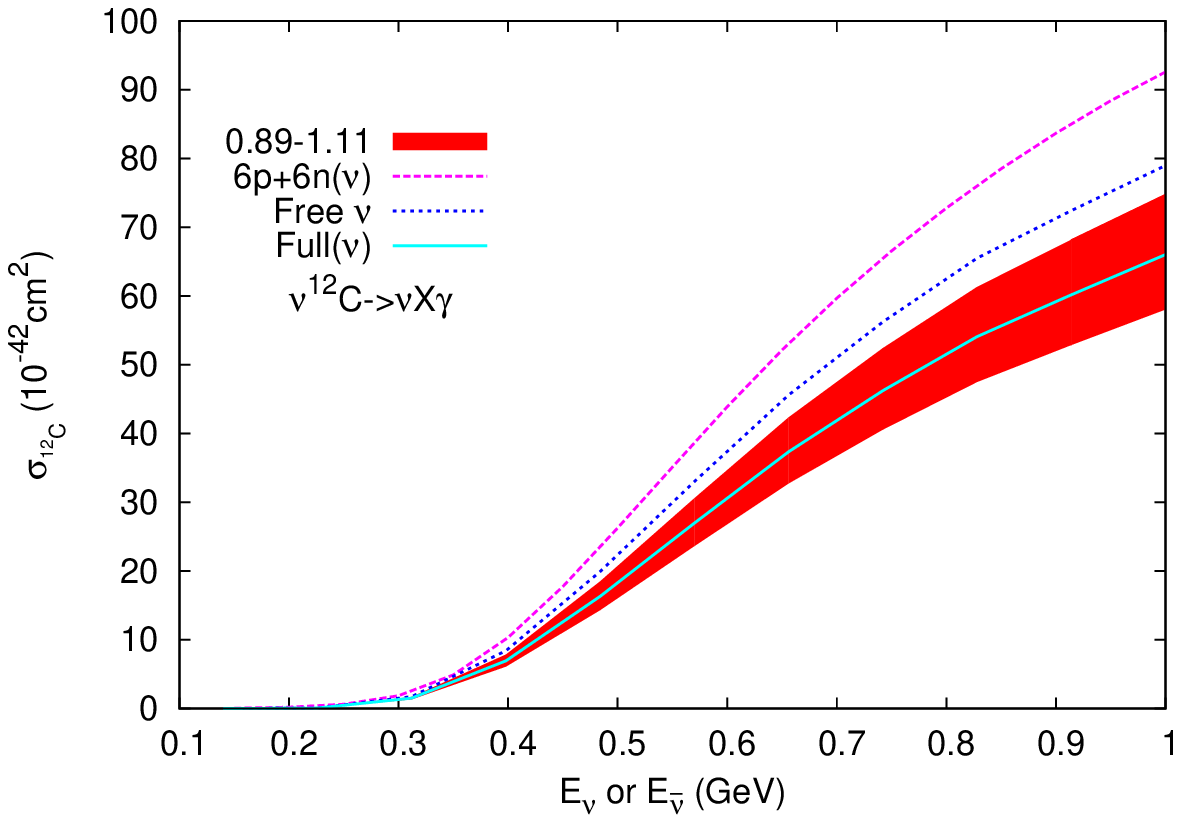}
\includegraphics[width=0.5\textwidth]{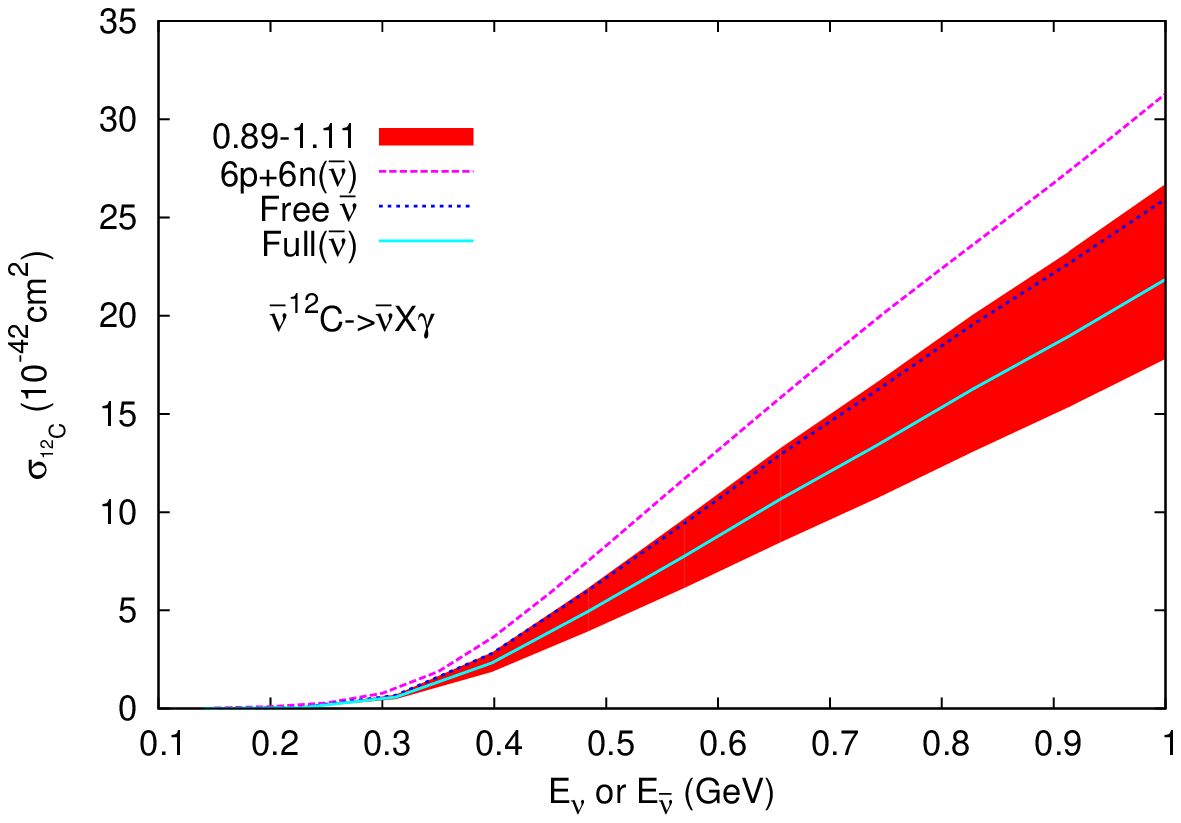}
  \caption{Integrated cross section for incoherent NC$\gamma$ on a $^{12}C$ target. The error band corresponds to the uncertainty in $C^A_5(0)=1 \pm 0.11$. }
\end{figure}

In Fig.~\ref{fig:resu_carbon} the cross section for reaction (\ref{eq:reac_incoh}) on $^{12}C$ is presented. The dashed lines are obtained by summing the contributions of 6 protons and 6 nucleons at rest, {\it i.e.} neglecting all nuclear effects. By taking into account Fermi motion and Pauli blocking, the cross section already goes down by more than 10~\%. With the full model, including the modification of the $\Delta$ resonance, the reduction is $\sim 30$~\%. A similar net effect is obtained in Ref.~\cite{Zhang:2012xn}, although the reduction quoted for the direct $\Delta$ mechanism (50~\%) is substantially larger than ours ($\sim 30$~\%).    

\begin{figure}[th]
\label{fig:events}
\includegraphics[width=0.5\textwidth]{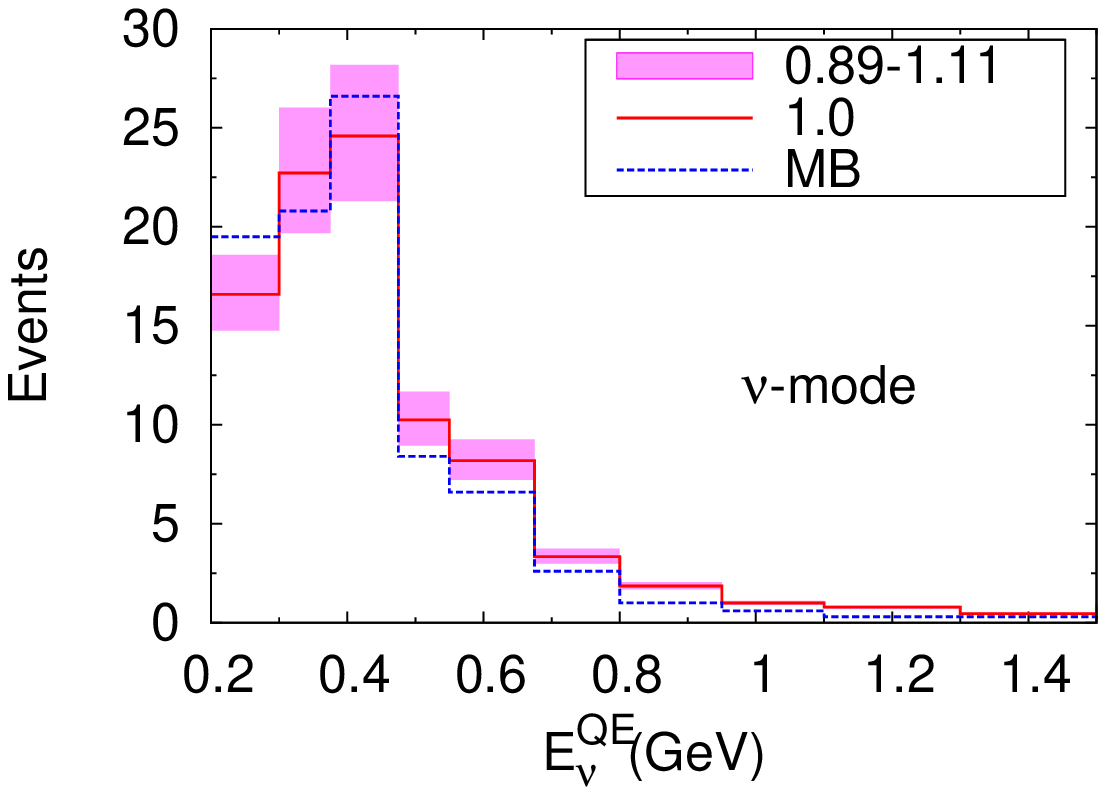}
\includegraphics[width=0.5\textwidth]{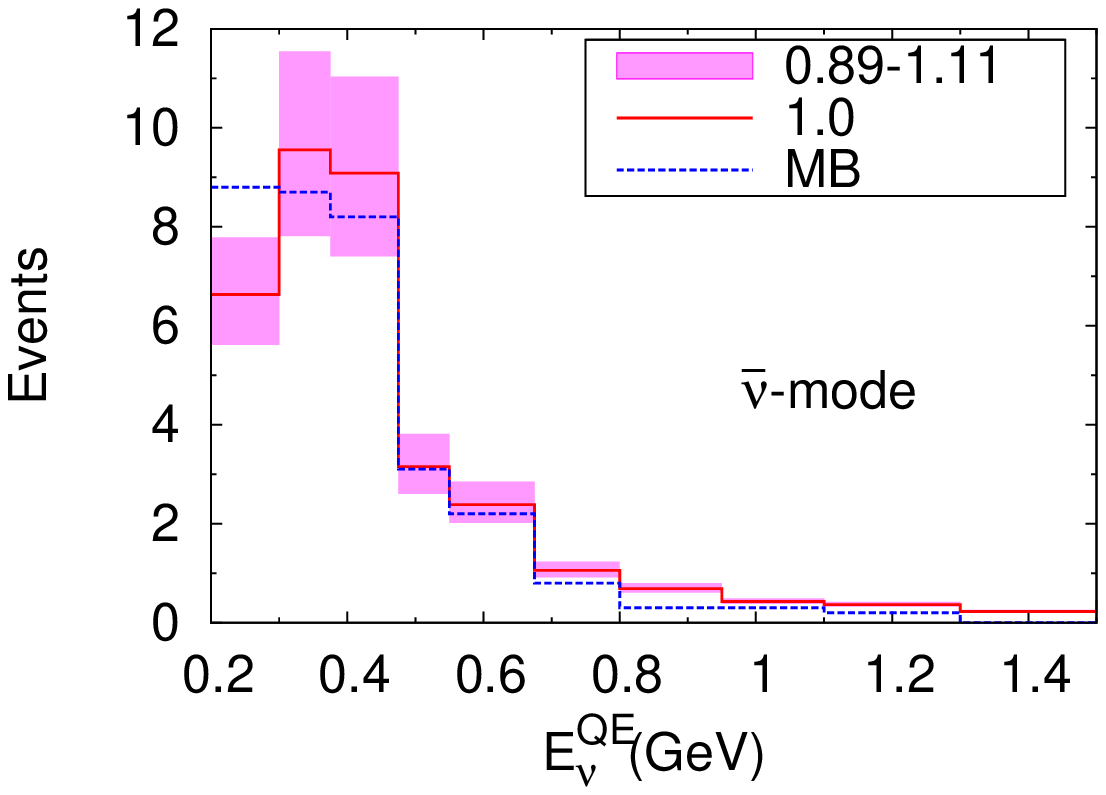}
  \caption{Distribution of $e$-like events at MiniBooNE as a function of the reconstructed (anti)neutrino energy. The solid line is the prediction of the present model while the dashed line is the MiniBooNE estimate based on NC$\pi^0$ measurements. The error band corresponds to the uncertainty in $C^A_5(0)=1 \pm 0.11$.}
\end{figure}
Finally, using the available information about the total number of protons on target ($6.46 \times 10^{20}$ in $\nu$ mode and $11.27 \times 10^{20}$ in $\bar{\nu}$ mode), the target mass (806 tons) and composition (CH$_2$) of the MiniBooNE detector~\cite{Aguilar-Arevalo:2013pmq}, as well as the (anti)neutrino flux determination~\cite{AguilarArevalo:2008yp} and the energy-dependent efficiencies for $\gamma$ detection~\cite{MiniBooNEweb} we have calculated the number of $e$-like events from photons according to our model (further details will be given elsewhere). The comparison to the MiniBooNE {\it in situ} estimate is shown in Fig.~\ref{fig:events} as a function of the (anti)neutrino energy reconstructed assuming charged-current quasielastic scattering on a bound nucleon at rest
\begin{equation}
E^{QE}_\nu = \frac{2(M -E_B)E_\gamma-\left(E_B^2-2M E_B\right)}{2\left[(M -E_B)-E_\gamma(1-cos\theta_\gamma) \right]},
\end{equation}
The $\nu_e$ and $\bar{\nu}_e$ components of the flux can be neglected as expected but the wrong sign ($\nu_\mu$ in $\bar{\nu}$ mode and viceversa) neutrinos yield a sizable contribution to the spectrum, in particular $\nu_\mu$ in $\bar{\nu}$ mode. We find that our results are in good agreement with the MiniBooNE determination. In spite of the quantitative differences in the models, we arrive at the same conclusion as in Ref.~\cite{Zhang:2012xn}, namely that NC$\gamma$ cannot explain the observed excess of $e$-like events at low $E_\nu^{QE}$.

%%%%%%%%%%%%%%%%%%%%%%%%%%%%%%%%%%%%%%%%%%%%%%%%
%% BACKMATTER
%%%%%%%%%%%%%%%%%%%%%%%%%%%%%%%%%%%%%%%%%%%%%%%%

\begin{theacknowledgments}
We thank T. Kaori, K. Mahn and G. Zeller for useful communications. Research supported by the Spanish Ministerio de Econom\'ia y Competitividad and European FEDER funds under Contracts FIS2011-28853-C02-01 and FIS2011-28853-C02-02, Generalitat Valenciana under Contract PROMETEO/2009/0090 and the EU Hadron-Physics3 project, Grant No. 283286.
\end{theacknowledgments}

\bibliographystyle{aipproc}   % if natbib is available

\bibliography{photonbiblio}

\end{document}